# Causality and Association: The Statistical and Legal Approaches

K. Mengersen, S. A. Moynihan and R. L. Tweedie[1]

*Abstract.* This paper discusses different needs and approaches to establishing "causation" that are relevant in legal cases involving statistical input based on epidemiological (or more generally observational or population-based) information.

We distinguish between three versions of "cause": the first involves negligence in providing or allowing exposure, the second involves "cause" as it is shown through a scientifically proved increased risk of an outcome from the exposure in a population, and the third considers "cause" as it might apply to an individual plaintiff based on the first two. The population-oriented "cause" is that commonly addressed by statisticians, and we propose a variation on the Bradford Hill approach to testing such causality in an observational framework, and discuss how such a systematic series of tests might be considered in a legal context.

We review some current legal approaches to using probabilistic statements, and link these with the scientific methodology as developed here. In particular, we provide an approach both to the idea of individual outcomes being caused on a balance of probabilities, and to the idea of material contribution to such outcomes.

Statistical terminology and legal usage of terms such as "proof on the balance of probabilities" or "causation" can easily become confused, largely due to similar language describing dissimilar concepts; we conclude, however, that a careful analysis can identify and separate those areas in which a legal decision alone is required and those areas in which scientific approaches are useful.

*Key words and phrases:* Statistical causation, epidemiological risk, relative risk, legal causation, tort, negligence, admissible evidence.

*K. Mengersen is Professor, Department of Statistics, Queensland University of Technology, Brisbane, Queensland 4001, Australia e-mail: k.mengersen@qut.edu.au. S. A. Moynihan is a Research Assistant, Faculties of Science and Law, Queensland University of Technology, Brisbane, Queensland 4001, Australia e-mail: s.moynihan@qut.edu.au. At the time of commencing this paper, Richard Tweedie was Professor, Department of Statistics, Colorado State University, Fort Collins, Colorado 80523, USA.*

[1]Richard Tweedie died before this paper was completed. His seminal contribution to the paper is acknowledged.

## 1. INTRODUCTION: THE ELEMENTS OF ASSOCIATION

Deciding whether relationships between outcomes and associated actions and events are causal lies, in quite different ways, at the hearts of both the legal system and the scientific process.

In the traditional sense, as developed in physics, for example, causal relationships are established in science by showing that a certain exposure is virtually inescapably linked to an outcome, through a process of repeatable experimental studies and through development of rigorous models for such linkage. A typical example of this traditional process is in the development of the laws of motion; every experiment shows how application of the same force consistently leads to the same motion, and we then uncover a "model" which describes how (if not why) the causal relationship "works."

No such experimentation is available for the development of explanatory models in traditional legal methods. A legal dispute as considered here is one which is typically resolved by considering general and specific actions and exposures involving a plaintiff who has suffered from some single rather than repeated outcome and deciding, through consideration of all known possible explanations, whether it has been shown "on the balance of probabilities" (in civil cases) that the defendant indeed "caused" the actions leading to the exposures which then further "caused" the outcome. In particular, we focus on legal cases that involve statistical input based on both population-level and individual-level information.

Here we have immediately and deliberately introduced a two-stage causal chain, from an *action* A alleged to have been taken by the defendant, to an *exposure* E alleged to be "caused" by the action, to an *outcome* O alleged to be "caused" by the exposure, operating at two levels, the population and the individual. The terms "action," "exposure" and "outcome" are used here as convenient and very widely covering terms, not intended to be interpreted in any way narrowly, and we will illustrate them with sundry examples.

One well-known example is the *Dalkon Shield cases* [60]. The Dalkon Shield is a contraceptive intrauterine device (IUD) which led to a string of complaints particularly regarding high pregnancy rates, infertility and pelvic inflammatory disease (PID). *Hawkinson v A. H. Robins Co., Inc.* [91] was an action brought by a class of women who had used the Dalkon Shield which they purchased from Robins. Robins had released a patient information brochure which claimed that women could safely wear the IUD for five years or longer. Case-control and cohort studies showed that the risks associated with the Dalkon Shield were substantially larger than those advertised by Robins. Robins was found to have negligently misrepresented the effectiveness and safety of its product.

Thus in a legal setting, in this case it had at least to be established that the defendant did take an action (say, failing to warn of hazards while selling the Dalkon Shield, or failing to test sufficiently for such hazards) responsible for the exposure (usage of the Dalkon Shield by the plaintiff) which involved a risk of harm. Then further it must be established that such an exposure did "cause" the outcome (such as PID) for which compensation is sought.

In a similar vein, in *Sindell v Abbott Laboratories* [102], the manufacturers of DES, a drug prescribed for the relief of morning sickness in pregnant women, were sued for causing a rare vaginal cancer in these women's daughters [20, 102]. Also, compensation was sought for skin and other health disorders by Vietnam War soldiers exposed to the defoliant agent Agent Orange [58, 93]; and workers in an asbestos mill sought compensation for health effects associated with exposure to asbestos [79]. These examples are discussed in more detail later.

The two-stage causal chain will be seen to hold for the individual plaintiff as well. For example, in *Ybarra v Spangard* [106], a surgical team was required to explain what had happened to cause an injury; in *Wilsher v Essex Health Authority* [105], medical defendants were charged with causing almost total blindness in a prematurely born baby by the administration of excess oxygen through misinserting a tube into a vein instead of an artery; and in *McGhee v National Coal Board* [97] the plaintiff was exposed to brick dust after work, suffered skin disease and accused the employer of being liable because shower facilities were not provided.

In this paper we consider the ways in which scientific and legal proof intertwine when the second of the "causal steps" described above, from exposure to outcome, involves scientific reasoning from population-based studies, in particular epidemiological studies. This is an important area for current debate, as in an increasing number of legal cases this step of the legal process involves an appeal to a population-based relationship: the *Dalkon Shield cases* [60], the effects of Agent Orange [58, 93] and the *DES case* [102] are major examples.

For population-based studies, as opposed to experimental studies, the general statistical concept of a causal relationship has a likeness to general causation in the legal system. Causation is asserted (as we develop in more detail in Section 3 and the Appendix) largely by observation of an otherwise unexplained association of the potential agent and the possible effect. Evidence from a number of sources is used to establish that results can be explained only by the potential agent, and not by other causes or by pure chance; and (at least until disputed by other evidence) the relationship is then accepted. For legal application, this general or population-based causation must be established at the level of exposure suffered by the plaintiff.

On the face of it, the legal concept of "proof on the balance of probabilities" and the population-based requirement of "statistical significance" appear to belong to the same traditions of proof: we accept the conclusion as being the most satisfactory possible even in the face of uncertainty or doubt. For practical purposes, in the scientific arena an association will not be considered to be confirmed at least until it is shown to be statistically significant. The legal concept is more lenient: if a court accepts a statement as proved on the balance of probabilities, it is accepted as certain for practical purposes even though it may have been found only to be more likely than not to be true.

Because of these similarities, when population-based studies are used in legal disputes, there is considerable possibility for confusion about the standards to be applied which should lead to a conclusion that causation has been "proved" in one or the other arena. One of our goals here is to clarify the differences in approaches, and to describe how the scientific ideas can in fact be used effectively in the legal context. Some of the legal questions and issues are stated by Peppin [48], Stapleton [61] and Price [49], among others.

However, there is one issue which lies at the heart of the difference between science and the law and this will be quite crucial in understanding their interaction. This is the fact that, for the law, there is in essence a single plaintiff (an individual, or a group of individuals in a class action) and the law has to decide on causation of outcome to each individual plaintiff; there must be consideration of the evidence relevant to the actual circumstances of the plaintiff. The population-studying scientist, in our examples usually an epidemiologist, seeks rather to decide whether overall the relationship affects the "population" as a whole, rather than whether any individual in the population is affected. Causality in a legal case must then allow for the transfer of such population-based arguments to individual circumstances, that is, from general to individual causation.

Within the court, then, at least two kinds of toxic tort causation must be proved: general or population-based causation (is the agent *capable* of causing the disease that the plaintiff suffers at the exposure suffered by the plaintiff?) through a preponderance of evidence, and specific or individual causation (did the agent cause *this* particular plaintiff's disease?).

Thus there will be a plaintiff who has suffered from an individual outcome which we denote $(O_{ind})$ and wishes to have it decided that the individual exposure, which we denote $(E_{ind})$ led to the individual outcome $(O_{ind})$; and equally centrally, that $(E_{ind})$ was due to an action $(A)$. Thus the relationship the plaintiff seeks to prove can be visualized as

$$(A) \to (E_{ind}) \to (O_{ind}).$$

The first link $(A) \to (E_{ind})$ requires the establishment not only of both the action and exposure, but also that performing the action $(A)$ involved negligence. Thus it is not so much the action $(A)$ that the plaintiff must prove. Rather, it must be proved that the defendant owed a duty of care to the plaintiff and that this duty was breached by the defendant taking the action $(A)$ and producing the exposure $(E_{ind})$, which may cause the plaintiff harm. It is only then that the question of culpability arises.

Often the court is focused more on testing this first link $(A) \to (E_{ind})$ and it may be quite clear that the second holds. As an extreme case example, it may not be in doubt that exposure to a dose of arsenic $(E)$ caused death $(O_{ind})$, and the question is clearly whether the exposure followed from the defendant's action. Equally, the role of science has often focused on the second link $(E) \to (O)$; in particular, epidemiology is concerned with whether the exposure in the population $(E_{pop})$ is linked with an outcome $(O_{pop})$ in that population. In the cases we consider, there will be doubts not only as to whether the individual outcome was caused by the exposure, but even whether for the general population this exposure causes the outcome.

Diagrammatically, then, the law wishes to assess whether any or all of the following set of implications (or causal chains) have been established:

$$(1) \quad \begin{array}{c} (E_{pop}) \to (O_{pop}) \\ \downarrow \\ (A) \to (E_{ind}) \to (O_{ind}). \end{array}$$

Our primary goal in this paper is to discuss the ways in which such relationships are established in a statistical and in a legal sense. This will be done in Sections 3 to 5. Prior to this, however, we will define more carefully the structure we have sketched above, and give more concrete examples of the types of links which may occur in a causal chain. These examples, taken both from existing legal cases on which judgment has been given, and from epidemiological studies not necessarily yet in legal dispute, are intended to illustrate a range of possible relationships and the issues that need to be given consideration.

Throughout this paper, we will take the view that causality must be proved, rather than "lack of causality" disproved: this is in accord with the statistical concept of commencing with a null hypothesis that there is no causality or even association, and it is attuned to the defendant's viewpoint in legal situations since the onus is placed on the plaintiff to establish that causal relationships really do exist. In Australia, legislation [75] provides that the plaintiff bears the onus of proving, on the balance of probabilities, that the defendant caused the harm. However, as Carver [5] explains, *Shorey v PT Ltd* [101] has provided that "the defendant still holds an evidential onus of proof to try to displace the inferences of causation supplied by evidence presented by the plaintiff." This shifting of the onus of proof often tends to occur where there are multiple possible actions or exposures, as discussed by Fleming [17]: for example, in *Ybarra v Spangard* [106] the surgical team was required to explain the actions that had caused an injury, although in *Wilsher* [105] the defendant was not required to prove that, say, low birth weight rather than the allegedly negligent treatment had caused the outcome of interest.

## 2. DEFINING THE ELEMENTS AND LINK TYPES IN A CAUSAL CHAIN

"Actions," "exposures" and "outcomes" can describe a wide variety of events. We will consider the situation in which there is a case brought by a plaintiff (an individual, or a class action brought by a group of supposedly similarly affected individuals) against a defendant (which may be a single entity or a group of entities defending together or separately). The complications caused by a multiplicity of plaintiffs or defendants are real ones in the legal system, but they do not affect the issues we discuss and we only briefly touch on them in Section 4.3.

We will find it simpler to work back from outcomes to exposures to actions.

### 2.1 Outcomes

We assume here that there is one outcome of interest which is an individual harm suffered by a plaintiff. Corresponding with the harm is loss and damage, without which there would be no liability and prima facie no action. Since we focus on situations in which there is a population-based relationship, we take an outcome to be usually an illness, disease or other personal injury, as with the onset of PID in the *Dalkon shield cases* [60] or cancer of the daughters in the *DES case* [102].

In our discussion, we typically assume that the occurrence of harm has been established. This may not be trivial, even when the outcome appears to be well defined, since some diseases are not easily diagnosed; for example, mild forms of PID may escape diagnosis; and some cases of asbestosis are not identified until after death. Although the court does need to be assured of the existence of the alleged outcome, interaction with statistical reasoning is limited at this stage, and we ignore this issue in what follows.

After determining that there is an outcome, a causal link must be established between the exposure and the outcome. Recent special cases have seen compensation awarded on the basis of increased risk. In essence, this implies a concept of virtual equivalence between $(E_{pop}) \to (O_{pop})$ and $(E_{ind}) \to (O_{ind})$ and it leaves it possible for the court to establish such equivalence merely by assuring that the individual is part of the relevant population for which the population risk is established. The interaction with statistical methodology would be almost paramount in such legal decisions.

### 2.2 Exposure, $(E_{pop}) \to (O_{pop})$ and $(E_{ind}) \to (O_{ind})$

"Exposure" is the name we give to the agent which is suspected of being causal, to which the population or individual is exposed.

The link between exposure and outcome can be dramatic and self-evident. In the case of a car accident, exposure is being hit by the vehicle, and the individual relationship $(E_{ind}) \to (O_{ind})$ needs no proof. But exposure in the situations we consider will often be far more subtle, and will involve the plaintiff claiming to be part of a more general exposed population: the users of the Dalkon shield, the workers in an asbestos mill, the soldiers exposed to Agent Orange in Vietnam.

There are three categories of possible relationship that we will differentiate. Note that we say "possible relationship" quite deliberately, since the whole goal is to determine whether the proposed relationship does actually hold. We give below a categorization of possible relationships which require somewhat different proofs of causation.

**R(0): Necessary and sufficient relationship:** In this situation, which we might denote $E \leftrightarrow O$, there is one possible exposure under consideration and we seek to show that it is necessary and sufficient for the outcome. This is the traditional meaning of "causal" in both legal and scientific arenas. Legislation in Australia [75] expressly requires that for causation to be established, the exposure must be a "necessary condition" of the harm (see Section 4). If sufficiency also holds, so that there is only one possible exposure, then assessing an individual relationship is more relevant than a population relationship.

The typical example is in cases of accidental or deliberate injury: the leg was broken ($O_{ind}$) because the car hit the plaintiff ($E_{ind}$), and conversely without the car hitting the plaintiff there would have been no broken leg; or the arsenic administration was both necessary and sufficient for the poisoning.

In population-based relationships it may be the case that the exposure is found to be sufficient [with, say, ($E_{ind}$) being asbestos exposure, and ($O_{pop}$) being occurrence of mesothelioma]; but it is much less likely that it is necessary. Pursuing the same example, even mesothelioma can be ideopathic (so that chance or background cases occur without asbestos exposure), if very rarely [53]. Our next two classifications delineate such situations.

**R(1): Relationships with a single identified exposure:** In this situation we have only one (in the legal context, potentially compensable) exposure of interest, denoted ($E^C$), or at least only one identified, which might have caused the outcome (O). There are, however, other known background cases of the outcome which occurred without exposure to $E^C$ and which were caused by chance or other (unidentified) exposures; we denote these collectively by $E^B$. This situation may be depicted as $(E^C, E^B) \to (O)$. In trying to prove, in the population, that $(E^C_{pop}) \to (O_{pop})$ is causal, we must show in essence that all of the outcomes are not just background, but that some are due to ($E^C$). Moreover, in anticipation of the legal case before us, we must show that this relationship holds at the level of exposure suffered by the plaintiff.

Typical examples of R(1) might include the *DES case* [102], or the relationship in a population between exposure to radiation and later occurrence of various cancers. In the first, the exposure ($E^C_{ind}$) is the ingestion of the drug DES and the outcome ($O_{ind}$) is appearance of the cancerous lesions of the vagina in the daughter after puberty. In appealing to the supporting causal link $(E^C_{pop}) \to (O_{pop})$, no other exposure was claimed to be explanatory for the outcome of cancer in the daughters, and it was found that the individual relationship was causal.

In population-based studies, exposure to high levels of radiation ($E^C_{pop}$) has been found to be associated with increased occurrences of particular cancers ($O_{pop}$) [78]. There is also a background level of cancer occurrences to be considered, but no other identified "nonbackground" exposure is identified as potentially causal.

The questions in R(1) usually concern the validity of the association $(E^C_{pop}) \to (O_{pop})$. Even though no other potential cause is identified, it is still necessary to prove that the observed association is not just a fortuitous juxtaposition of occurrences of both exposure and outcome that has been observed, or that some other confounding factor is not responsible for the observation, as we discuss in Section 2.4.

**R(2): Relationships with several identified exposures:** "Background exposure" is a relatively unsatisfactory portmanteau phrase for exposures which might be explanatory of an outcome, and seeking to find other explanations is common to both the law (because such alternatives might provide a more plausible defense than mere chance or background) and to science (in which finding such potential explanations is usually the crux of ongoing research).

Here we have several identified contributory exposures $(E^C, \{E^j\}, E^B)$ which may be responsible for (O): $E^C$ is the specific exposure for which compensation is sought; $\{E^j\}$ is a set of other possible identified exposures and $E^B$ is background exposure.

In proving, even in the population, that $(E^C_{pop}) \to (O_{pop})$, we have to ensure now that the outcomes are neither just background nor all due to the alternative exposures. It has certainly been seen as a harder decision for the courts to rule in individual cases that the compensable exposure has been causal if other exposures can be positively identified which might have caused the outcome. This is particularly the case if the other exposures are noncompensable.

It is instructive to consider two examples which indicate the potential for different conclusions when on the face of it the framework is similar.

In *Wilsher v Essex Health Authority* [105], the exposure ($E_{ind}^C$), the excess oxygen given at the birth, was not in dispute. However, the fact that there were several other identified potential causes of the outcome ($O_{ind}$) of almost total blindness in the baby (including low birth weight, apnea and number of transfusions) seemed to make a court determination difficult. This was exacerbated in this case because the association between ($E_{pop}^C$) → ($O_{pop}$) varies according to exposure levels and because the status of each of the other population relations ($E_{pop}^j$) → ($O_{pop}$) is not clear-cut.

On the other hand, in *McGhee v National Coal Board* [97] the plaintiff argued that he was exposed to brick dust after work ($E_{ind}^C$) as his employer did not provide shower facilities. This brick dust exposure allegedly caused dermatitis ($O_{ind}$), and the general causation or population link ($E_{pop}^C$) → ($O_{pop}$) appears undisputed. Although he was also exposed to brick dust at work, which is an exposure in the set $\{E_{ind}^j\}$ associated with ($O_{ind}$), in this case it was found that the link ($E_{ind}^C$) → ($O_{ind}$) was proven and that it was of sufficient contribution relative to $\{E_{ind}^j\}$ to be compensated.

### 2.3 Actions

Actions are those things done or left undone by the defendant which are alleged to have caused the exposure. It is often not the action per se that the plaintiff must prove, but that the defendant breached a duty of care to the plaintiff, which caused the plaintiff harm.

These may be obvious actions, such as driving the vehicle which struck the plaintiff, or firing the bullet which killed the victim. They may be slightly less obvious or more disputable actions, such as allowing asbestos fibers to be free in the workplace as in [86] or misinserting a tube into a vein rather than an artery leading to excess oxygen as in *Wilsher* [105]. There may also be actions which are more properly termed inactions: the failure to adequately test products, or the failure to advertise potential harmful effects, both of which are relevant to PID occurrence associated with use of the Dalkon Shield [60] and the failure to provide showers for removing brick dust after work, as in *McGhee* [97].

Within population-based cases, one situation in which the role of statistical analysis may become increasingly important is where there is an "indeterminate defendant" [17]. Even if exposure is proven and the association ($E_{ind}^C$) → ($O_{ind}$) is accepted, it may be difficult to determine which one of many potential defendants caused an exposure in a particular plaintiff. Here, then, there is a set of actions ($A^C, \{A^j\}$) leading to ($E_{ind}^C$) but the particular compensable action $A^C$ is unidentifiable.

This occurred, for example, in the allegation of exposure to DES in *Sindell v Abbott Laboratories* [102]. Here the actual manufacturer of the drug ingested by an individual plaintiff's mother was not known and could have been one of a number of different companies: five companies were sued as potentially compensable. In this case a "market share" approach to allocating compensation for the step (A) → ($E_{pop}$) was adopted and individual plaintiffs did not have to prove that a specific defendant was responsible for (A) → ($E_{ind}$). The courts have, however, begun to move away from this market share approach; this is considered in more detail in Section 4.3.

So there is a role for scientific proof in showing that actions caused exposures, especially through forensic sciences and perhaps through population-based arguments, but in general it is not the concern of our thinking here.

### 2.4 Describing the Causal Links

Once the events O, E and A have been established, we can inspect the links between them and make, in particular, two assessments: whether the link is direct or indirect, and how to describe any interaction between multiple events. As illustration, here we focus on the link between an exposure of interest, $E^C$, and a single outcome O. Three types of links between $E^C$ and O can be identified.

*No confounding of* $E^C$. In this situation $E^C$ is believed to be directly related to O, with no other identified exposures which might mitigate the association. The cases R(0) and R(1) may be described in this way.

*Indirect link through a confounding exposure.* In this situation we find that although an association between $E^C$ and O is observed, there is no actual causal relation. Rather, $E^C$ impacts on, or is closely associated with, another exposure E which is causing O, in such a way that $E^C$ is rendered noncausal.

An excellent example of this in the epidemiological context is the relationship between poppers and AIDS. In the early part of the study of AIDS, population-based studies showed that users of "poppers" (amyl nitrate) were associated with AIDS occurrence, or with being HIV-positive, more than were nonusers of poppers. We can then ask whether this is a causal relationship: do poppers, as an exposure (E), cause AIDS as an outcome (O)? Ultimately it was established that was not the case: poppers were associated closely with certain sexual activities, which were satisfactorily shown to be the causal agent in the spread of the virus.

*Combined exposures, contributing directly and indirectly to outcome.* The third situation which may be distinguished, and which occurs often under R(2), is one in which there are interconnections between the various exposures and both the potentially compensable exposure $E^C$ and another exposure E (or other exposures $\{E^j\}$) are all contributing causally to an outcome.

This is by far the most difficult to deal with. Continuing the AIDS example, consider a population exposed both to homosexual practices and to blood transfusions. Both of these are known to lead to increased incidence of AIDS. Here there is no reason to expect that being homosexual increases the predisposition to contract AIDS from blood transfusions or vice versa. The increased incidence should therefore be essentially the sum of the increased incidences from the two separate exposures considered independently.

We might express this link in which the two exposures independently affect the outcome in a population by an *additive* model depicted diagrammatically as

$$[E^C_{\text{pop}} \to O_{\text{pop}}] + [E^1_{\text{pop}} \to O_{\text{pop}}]$$

with $E^C$ representing blood transfusions, say, and $E^1$ homosexuality.

As a second example, we take studies of the association between asbestos and lung cancer (and, of course, other diseases such as mesothelioma). At the Wittenoom mine in Western Australia, exposure to airborne asbestos fiber (crocidolite) appears to have been widespread ($E^C_{\text{pop}}$). Studies of the population of exposed miners indicate an increase in the incidence of lung cancer, when compared to other unexposed persons [1]. Industrial exposure to amosite and chrysotile asbestos dust has been studied in England and the United States [1]. However, the majority of asbestos workers are also active smokers, and the increased incidence of lung cancer in active smokers is well documented [13]. Hence again we have a situation in which the relationship of asbestos to lung cancer ($O_{\text{pop}}$) is confounded with the relationship of smoking to lung cancer as an alternative exposure ($E^1_{\text{pop}}$).

As a further example we note that in the studies of IUD and PID occurrence, there is confounding in an unexpected way: it appears that the use of oral contraceptives is likely to be protective for PID occurrence, so that if (as is usually the case) there are oral contraceptors in the control or unexposed group in a study, then this group will show lower than normal PID occurrence, which results in the exposed group appearing to have a relatively higher than normal PID occurrence rate.

All of these possibilities might be depicted through a synergistic model, in which the various exposures interact and contribute more (or sometimes less) excess incidence of the outcome than would be expected from the individual exposures alone. Diagrammatically we may write this as

$$(2) \qquad (E^C_{\text{pop}} + E^1_{\text{pop}} + E^C_{\text{pop}} \bullet E^1_{\text{pop}}) \to O_{\text{pop}}$$

with $E^C_{\text{pop}} \bullet E^1_{\text{pop}}$ indicating the interactive effect of $E^C$ and $E^1$.

A variant on this situation occurs in the assessment of causal relationships between the use of an IUD and outcomes of PID and infertility. The use of the Dalkon Shield ($E^C_{\text{pop}}$) appears to be associated directly with increased incidence of infertility ($O_{\text{pop}}$). It is also directly associated with PID ($O^1_{\text{pop}}$), which is itself directly and causally linked to infertility. Thus we have both a direct and an indirect link between ($E^C_{\text{pop}}$) and ($O_{\text{pop}}$).

We discuss briefly in Section 4 the way in which the courts might deal with such multiple causes, once established; this, however, is their role and not the role of statisticians or economists, as has been well expressed by Robins and Greenland [51]. Rather, our emphasis in this paper is on the way in which the two disciplines manage to prove that any such relationships are established at all.

### 2.5 Testing for Causation

A legal two-limb test for causation using the results of an epidemiological study was described in *Seltsam* [100]:

1. *General Causation—Is the exposure, more probably than not, capable of causing or contributing to the outcome in the population?*
2. *Specific Causation—Was the outcome in the individual case, more probably than not, caused or contributed to by the exposure the individual was subjected to?*

This legal test is almost the equivalent to the two-step scientific test for causation defined in Section 1.

1. Has a "scientific" causal relationship ($E_{pop}$) → ($O_{pop}$) been established in the (relevant) population?
2. *On the balance of probabilities, was the plaintiff's individual outcome caused by the exposure?*

We now consider the establishment of these two steps separately.

## 3. ESTABLISHING GENERAL OR POPULATION-BASED CAUSATION

We restrict ourselves here to situations involving observational studies on human populations, that is, to the methodology one might employ to prove that the relationship ($E_{pop}$) → ($O_{pop}$) is actually causal.

Moreover, for focus we will concentrate on epidemiological studies of the effects of exposure to allegedly or potentially harmful substances, so that there is a compensable outcome. Thus we are specifically thinking of claims against defendants, and not forming public policy. For example, the former requires establishment of causality before a defendant is ordered to pay compensation to individual plaintiffs, and consequently more detailed consideration of the types of links between ($E_{pop}$) and ($O_{pop}$); the latter may require a more lenient demonstration of association ($E_{pop}$ → $O_{pop}$), sufficient to support the adoption of a policy of prudent avoidance of the exposure.

Nevertheless, policy considerations have an integral role in determining causation in a negligence action. This was first established in Australia by the High Court in *March v Stramare* [96] where it laid down the commonsense and experience test. The recent civil liability reform in Australia has placed a greater emphasis upon the analysis of policy and commonsense in determining in tort cases [8]. Because of this, there can be no strict universal level of contribution indicative of causation since this can change according to the facts and circumstances of the case. Nevertheless, for there to be causation in any given case, the principles set out in this section and in Section 4 must be satisfied [53].

Spigelman [100], paragraph 183, found that the determination of whether the evidence is capable of bearing the inference is for the courts to decide using a "commonsense approach," and commented on what would be commonsense in terms of examining epidemiological evidence:

> "*the proposition that the stronger the association the lower the probability that it would occur without a causal relationship, is a commonsense proposition which a court will readily accept. The same is true of the proposition that inconsistency of results undermines an inference of causation.*"—at paragraph 147.

It is thus incumbent on the scientist to clearly establish the epidemiological support for a causal argument prior to a court's consideration of this evidence.

### 3.1 Philosophical Theories and Pragmatic Tests of Causation

There exists a substantial philosophical theory of causation. Much of the philosophical writing on this topic is concerned primarily with the problem of establishing cause with *certitude* (see, e.g., [55] for several papers on the writings of Hume and Popper in relevant areas; also see [26]).

We will accept without any great surprise or concern that in both the legal and applied statistics arenas an established cause is always subject to falsification: new evidence, new experiments can overturn previously accepted decisions or theories, and even Newton's Laws of Motion, perhaps as well established as any causal association can be, were subject to substantial modification by Einstein. We are concerned, rather, with how, on the facts available at a given time, one might assert that causation is satisfactorily shown.

We will argue the view that, both for practicing scientists and for lawyers, proof of a causal relationship is:

(a) provided by passing a number of tests of a general nature, as delineated in Section 3.4 and the Appendix,
(b) susceptible of qualification, with "strong" or "weak" proof being reasonable concepts, based

on the way in which the tests in (a) are passed, and also on a knowledge of the importance of the assumptions made (often implicitly) in reaching conclusions,

(c) susceptible of later reversal or falsification, in the light of new data or concepts being accepted.

It is this last point (c) which allows lawyers and scientists to live comfortably with operational ideas of causality somewhat outside the philosophical attitudes of Hume or Popper. The falsifiability, refutability or testability of a scientific method is a key criterion of the method's scientific status and also of its legal status since the decision of *Daubert* [85], which cited Popper's approach in the opinion (see Section 3.5 for more detail about the legal acceptance of a method).

One cause of (c) in a current legal context is the limitations period: the scientific knowledge may not yet be complete, but still the Court must make a decision regardless of subsequent developments [21]. In the adversarial process, the duty is on the parties, not the Court, to produce the evidence. The Court is then bound to make a decision on the basis of the evidence submitted. As Gastwirth [21] explains, even if the decision is made in the absence of important studies or literature or if studies arise after the decision which contradict the decision, the legal decision is not necessarily wrong. If the Court gives due consideration to each piece of evidence and justifies the reliance placed on each piece of evidence, then the Court will have fulfilled its duties [21].

It is worth stressing that the second point (b) may not be fully realized by the courts as applying to scientific matters. Although lawyers would accept all of the above points as true of legal reasoning, statistical or epidemiological evidence may be admitted as "a question of scientific fact"; and there then may be an over-reliance on the certainty of scientific conclusions beyond what the scientific community itself would expect.

In practice, however, it is the statement as in (a) of an acceptable and explicitly agreed set of tests required for a working "proof" that seems to be the crux of allowing the legal system to use scientific assertions of causality. It is from such a set of tests, whether explicitly laid out or implicitly assumed, that we may be able to move from the existence of empirically observed "positive association" (without which, of course, we would rarely have any case to argue concerning causality) to an agreed position that the association is causal. Before detailing such a set of tests, we must discuss this particular aspect: the identification of a measure of association between (E) and (O) in the population.

### 3.2 Establishing Cause Through Raised Population Risk

Much of the discussion of Hume and Popper in the philosophical literature also relates essentially to the situation in which there is (potentially) a *necessary* or a sole or sufficient cause; that is, when R(0) holds.

Although it is also traditional in law to consider situations where $(E_{ind}) \to (O_{ind})$ similarly follows because of a necessary causal relationship such as R(0) ("the arsenic was swallowed and therefore the victim died"), in the cases of relevance to us we do not have this absolute causation. Instead, we have population-based outcomes, often rare, which can be expressed as *relative risks* and which usually appear to be raised above the norm by the exposure. Our concerns are thus typically with the cases R(1) in which there are background occurrences of the outcome (as in the *DES case*) or R(2) with a number of potential exposures which might have caused the outcome (as in the *Dalkon Shield* or *Wilsher cases*).

It is important to describe this carefully. Even in situations with close relationships between exposure and outcome, there are some ideopathic cases; and in situations in which there is a weaker association, such as the occurrence of cancers which may or may not be caused by radiation, there will be many population outcomes occurring without any radiation exposure. To be specific, if we have (say) 1000 people totally unexposed to $E_{pop}$, some number $M_{unexp}$ will contract the outcome; and if we have (say) 1000 people, all exposed to $E_{pop}$, some other number $M_{exp}$ will contract the outcome. We would then measure the relative risk in this population by RR = (expected rate of exposure per exposed person)/(expected rate of outcome per unexposed person) = $M_{exp}/M_{unexp}$. For example, if in observing a population we find there are 25 lung cancers annually per 1000 persons exposed to asbestos, compared to 5 lung cancers annually per 1000 persons not exposed to asbestos, then there will be a value for RR of 5.0. This means that on average, an exposed individual is five times as likely as an unexposed individual to suffer the outcome.

In case-control studies of rare outcomes the relative risk is adequately approximated by the odds ratio, expressed as the ratio of expected rates of exposure for exposed (case) and unexposed (control) subjects. Other measures, such as excess risk, standardized differences and mortality ratios may also be used. Complementary descriptions are also available, such as the "probability that a case is due to the exposure" $\text{PDE} = (\text{RR} - 1)/\text{RR}$ [20]. Thus a RR of 5 in the above scenario is equivalent to a probability of 0.8 that "the case is due to the exposure."

If $\text{RR} = 1$ (or $\text{OR} = 1$ or $\text{PDE} = 0$), then clearly there is no relationship between $E_{\text{pop}}$ and $O_{\text{pop}}$. If we can prove that $\text{RR} > 1$, then at least some of the outcomes in the population are associated with $E_{\text{pop}}$. However, it is important to remind ourselves that these are primarily measures of association (despite the wording "is due to" in the definition of PDE) and, as we discuss in Section 3.3, it may take further steps to assert general or population-based causation.

As the Courts determine civil matters on the balance of probabilities, a $\text{RR} > 2.0$ or $\text{PDE} > 0.5$ is as a rule of thumb indicative of causation. Proof on the balance of probabilities, as required by civil courts, is sometimes called the "51% rule" [17], which states that if there is more than a 50% chance of the outcome being causal, it is to be treated as causal. If we accept a causal link in the population, the apparent interpretation of a relative risk above 2.0 suggests that more than 50% of the outcomes ($O_{\text{pop}}$) are caused by the exposure of interest ($E_{\text{pop}}$) and less than 50% are due to other causes.

This rule has been the subject of considerable discussion and criticism. For example, since the rule is based on an estimate (rather than the true RR), it has been argued that it may be more appropriate to use the corresponding lower 95% confidence level (LCL) and require the more stringent test that this LCL must be above 2.0. In a different vein, Greenland [22] argues that it is important to understand the nature of the relationship between the exposure and the outcome before making such a rule: at the point at which the probability exceeds 50%, the exposure level may be well below that at which the incidence of disease is doubled. Maldonado and Greenland [42] and Greenland and Robins [23] point to the phenomenon of "accelerated outcome" (some of the outcomes in the exposed group would have occurred later anyway had the individual not been exposed) in biasing the relative risk in favor of the defendant. Although an exposure may accelerate the contraction of a disease, it does not necessarily cause the disease, because the individual was "doomed," the contraction of the disease was inevitable.

The $\text{RR} > 2$ rule has also been discussed in light of bias arising from shortcomings in the epidemiological study. Carruth and Goldstein [4] argue that overestimation might occur through the "incomplete accrual" problem, in which statistically significant associations observed in large epidemiological studies are published before all the cases of disease have accrued. Alternatively, underestimation might occur through the "healthy worker effect" if the epidemiological study targets a work force that is generally healthier than the general population. Moreover, if a raised RR is observed, "remedial action" may be taken to reduce the exposure, leading to lower RR estimates in subsequent studies.

It is apparent that using an overall relative risk of 2.0, say, as a cutoff point for all individuals in a given population is likely to be unpalatable in both scientific and legal contexts. However, in practice such a risk is, in almost all cases, a mixture of risks: there will be subgroups of individuals whose relative risk is well above (and well below) the average. By finding where the individual plaintiff lies in this mixture, it may be generally much easier to ensure that those almost certainly suffering because of the exposure are recompensed and those almost certainly suffering due to background or other causes are not recompensed. If we are able to use the actual exposure level and other factors to identify a relevant subgroup to which the individual belongs, then it is clear that the harshness of the "balance of probabilities" rule as an all-or-nothing approach is very substantially softened.

Of course, caution should be exercised in applying the 51% rule in other contexts. For example, from a public policy perspective, it would seem unreasonable to allow society to be subjected to an exposure that produced a relative risk within the range 1.5 to 1.9 [20, 78]. As discussed at the start of Section 3, recent reforms in the law of negligence have given courts the ability to use policy as a determinative factor when considering causation, so that the defendant may be liable for negligence even though the $\text{RR} > 2$ rule is failed and scientific proof of causality is not established.

### 3.3 Establishing Cause in a Relevant Population

The first of the subtests in the two-step procedures defined in Section 2.5 is satisfied if the epidemiological study proves that, more probably than not, there is an association between the exposure and the outcome in the population. If the exposure, whether due to negligent or other acts on the part of the defendant, has not been satisfactorily shown to be causally linked to the claimed outcome in the population and exposure category to which an individual belongs, then there is no case to argue based on population data.

As discussed above, such a debate may well involve the vital and often poorly handled question of using a *relevant* estimate of relative risk for the specific individual in question.

Relative risks are designed to quantify the excess outcomes that can meaningfully be ascribed to the exposure (E) within a whole population. But if there are other risk factors for (O) that are known, then the contribution of (E) should be measured for the relevant subgroups of populations or adjustment made for the other factors in the analysis. Such risk factors may be quite neutral from a legal point of view. The most obvious is age: exposure of an old person to a drug may change risk levels in quite different ways than does exposure of a younger person. They may be more contentious: asbestos exposure for smokers raises relative risk of lung cancer far more than for nonsmokers, and for any plaintiff the correct joint exposure should clearly be determined if possible.

The importance of identifying a suitable reference population is underlined by the various paradoxes that may arise if appropriate account is not taken of confounders. For example, Simpson's paradox [45] shows that a smaller relative risk estimate may be observed after the combination of groups with higher individual relative risks. Slud and Byar [59] provide further discussion of this phenomenon.

Adjustment for special circumstances is often done in a general way for whole populations. But here we face the question of assessing the increased risk of an outcome (O) from exposure (E) for a single plaintiff, and we may know a considerable amount about other risk factors for this individual or group of individuals.

Thus we must either:

(a) estimate the risk *relative to a population of persons similar with respect to other risk factors*, or

(b) *assume* that those other factors are effectively neutral, that is, that the relative risk appropriate to the plaintiff is the same as for any other person in the population.

The latter will often be demonstrably implausible to some degree, but it is in general a scientific impossibility to estimate risk for a very detailed subgroup of the population (e.g., those women of age 40 with three life-time sexual partners, use of an IUD for 6 months at the age of 27, and a family history of infertility; or those men aged 70, with 4 months in an asbestos mill, but not in the "dusty" part, and who smoked for 15 years but only using a filter, and who had worked as plumbers with exposure to various documented carcinogens for 30 years). Even at a less detailed level there may be limited information. For example, although there is substantial data about the relative risks of PID associated with IUD use, promiscuity (which can be a confounding and possibly causal factor) can be poorly documented, so that the relevant subpopulation may be difficult to define for an individual; or cross-classification of smoking and asbestos may not be precise, since asbestos could typically be very poorly measured.

It is understood by the courts that the *level of exposure*, both of the plaintiff and of the reference population, is not always available and while such would be beneficial, it is not necessary to demonstrate a substance is toxic to humans given substantial exposure [104]. If the plaintiff comes from a population with high exposure and a precisely estimated relative risk of 20.0, this may well indicate a convincingly causal association, whereas a plaintiff with low exposure coming from a population with a relative risk of 1.01 will have less grounds for such a conclusion.

This then raises the problem of measurement in the population and for the individual. Evaluation of the effects of asbestos provides a good example of many of these problems. For example, it may not be in contention that asbestos in the air results in some level of exposure to asbestos in the lungs of workers. But measuring the level of exposure in a population is notoriously difficult; see, for example, [34, 71], where different methods of measurement led to different conclusions on the same population. This is then exacerbated because we rarely have direct measurement of exposure in the lungs in the individual plaintiff, so the level of exposure of the plaintiff in a given case may be a matter of statistical estimation.

Of course, population studies cannot always be conducted at the level of detail desired to give results for a population relevant to a specific plaintiff, and if they are, numbers in the studies are almost bound to be so small that other aspects such as lack of statistical significance, accuracy of data and representativeness of the sample tend to render results open to criticism.

These considerations all highlight the need for accurate and detailed data, which is obviously well recognized by both scientists and lawyers. While both parties would accept that better quality data provide more reliable conclusions, courts may find difficulty in using epidemiological results since poor quality data, even though it may be indicative in the scientific arena of certain relationships, may be quite unconvincing under the scrutiny of the adversarial system.

Despite all of these difficulties, and even if a completely relevant population cannot be identified, statisticians can generally provide the court with some partial answers based on population data: the relative contribution of other risk factors (e.g., whether the number of sexual partners is important to the outcome of PID and if so, to what degree), relevant population subgroups (e.g., cohorts in which the RR is reasonably stable), a model which describes the individual outcome in terms of a range of factors or quantification of the effect of poor quality studies.

It rests with the court to determine, with guidance from the scientists, just how the individual fits into the relevant population, and consequently how much compromise can be borne in the assessment of the individual plaintiff.

### 3.4 A Test-Based Framework for Scientific Proof of General Causation

The approach we outline here brings together pragmatic tests through which, if passed to a "satisfactory" degree, causality may be deemed proven: if failed, then causality is still unproven. The use of a test-based approach seems to fit much of the thinking in both the scientific and the legal arenas, and perhaps forms one area of agreement between them when so many other aspects are different.

One of the most commonly used set of tests in the epidemiological context is generally attributed to Bradford Hill [30], which appears with modifications in many places (see, e.g., [54, 55]). Other recent approaches in the area of clinical medicine include Breslow and Day [3] and Chalmers [6].

Based on this literature, we list below ten tests which are relevant to asserting general or population-based causality. We spell these out in much more detail in the Appendix.

TEST 1 (Existence of mechanism). *Is the proposed association explained by a biologically plausible mechanism?*

This is not always a reasonable question given current scientific knowledge, and if failed we might turn to

TEST 2 (Analogous relationships). *Is the proposed association analogous to some other accepted causal association?*

There must be at least some reason for believing that the exposure should give the outcome.

TEST 3 (Temporality). *Does the exposure precede the outcome?*

This is obvious at first, but in many epidemiological situations the latency of the disease must be taken into account and the time of onset of disease may be indeterminate.

TEST 4 (Validity of data). *Are the data, on which the conclusion is based, valid?*

Mistakes, systematic biases and other errors in the study design, data collection and data entry must be ruled out.

TEST 5 (Strength of association). *Is the observed association strong, as measured, for example, by a RR substantially greater than 1.0?*

Regardless of whether it supports a real negative effect or is simply ambivalent about the effect due to small sample size, a RR less than 1.0 cannot lend support to a claim of causation.

TEST 6 (Lack of confounders). *Are there other aspects of the study group that might explain the observed association?*

As described in Section 2.4, there are often many other factors that must be ruled out as potential confounders or explanations of an observed association.

TEST 7 (Consistency of association). *Is the association consistently found over a number of studies?*

This conforms to the usual scientific principle of repeatability and provides robustness to the causal claim. There are reasons why such consistency may not be found, but these must be clearly established if this test is to be discounted.

TEST 8 (Statistical significance). *Is the observed association statistically significant?*

This is central. In scientific reasoning the probability that an observed positive association is due to chance fluctuation or "background causes" must be satisfactorily small.

TEST 9 (Dose-response relationship). *Is there an increase in magnitude of outcome from an increasing level of exposure?*

This is particularly important for assessing subgroups within a population and for considering individual causation, in particular evaluating the effect of exposure at the level experienced by the individual.

TEST 10 (Validity of logic). *Is the conclusion actually justified by the data and analysis presented?*

This appears obvious, but optimistic (or pessimistic) generalizations and extrapolations of the results of an analysis are common.

Before a strong case for $(E_{pop}) \to (O_{pop})$ can be made, we would argue that each of these tests needs to be considered. In some cases there may be a reasonable explanation why they might fail (as is the case in many of the examples we have proposed; see the Appendix for more detail); but unless they are considered appropriately, we believe that it is not possible to make a serious case that, in the population, $(E_{pop}) \to (O_{pop})$ has been proven.

## 3.5 Admissibility of Epidemiological Studies into Evidence

We turn now to the question of when epidemiological studies and the corresponding explanations by epidemiologists are admissible as evidence. Typically, such a study can be admitted into evidence as an expert opinion. Preconditions differ among the United States, United Kingdom and Australia, but there are some common themes among these jurisdictions.

All of the jurisdictions require that for any evidence to be admissible, it must be relevant to the issues contested in the litigation. In Australia, s55 of the *Evidence Act* (Cth) [76] provides that evidence is relevant if it

> "*could rationally affect (directly or indirectly) the assessment of the probability of the existence of a fact in issue in the proceeding.*"

In the United States rule 702 of the Federal Rules of Evidence [77] provides that expert evidence must be relevant such that it has a valid scientific connection to the pertinent inquiry. An epidemiological study will satisfy the test of relevance given it examines the relationship between the particular exposure and the particular outcome that is the basis for the case.

The more important issue is whether the methods used in the epidemiological study have an underlying foundation capable of forming the basis for an expert opinion. The United States has given more consideration to this issue than the other jurisdictions. Rule 702 of the Federal Rules of Evidence [77] provides that there must be reliability in the methods used by the expert and reliability in the application of these methods to the facts in the case. *Daubert* [85], *Joiner* [90] and *Kumho Tires* [95] are known as the "trilogy" and are the leading authorities on this point of reliability of the expert's opinion.

*Daubert* [85] provided that the test of reliability of expert opinion is a "flexible one" and that the trial court is to act as "gatekeeper" in applying this test. However, *Daubert* emphasized that the main concern is the reliability of the scientific methods, not the conclusions that they draw. The court then set out a nonexhaustive list of four factors to consider when determining whether the scientific method used by the expert is reliable: whether the expert's method has been tested; whether the expert's method was the subject of peer review or publication; whether the expert's method has a known or potential rate of error and standards controlling the method's operation; and whether the expert's method has general acceptance within the relevant scientific community.

The later decision of *Kumho Tires* [95] provided that these four factors may be given consideration by the court, but they are by no means strict requirements. Rather, the primary consideration for the court is whether the expert's opinion is of "the same level of intellectual rigor" as the expert would employ outside the courtroom when working in his or her relevant discipline.

The object of the "intellectual rigor test" is to prevent the expert reaching an opinion that is tailored for the purposes of litigation. As pointed out in *Joiner* [90], the expert must account for "how and why" he or she reached his/her opinion [2]. As Berger [2] writes, experts must show that their conclusions were reached by methods that are consistent with how their colleagues would proceed were they presented with the same facts and issues. So while a scientific expert can depart from canonical methods, he or she must then show grounds consistent with the methods of the scientific community that support the departure [83].

In Australia, for an expert opinion to be admissible it must be derived from methods that are accepted by the field of expertise [29]. While the High Court itself has not ruled on this, it does appear likely that the courts will use the criterion of general acceptance within the professional community, which has been rejected in the United States [18]. The court does, however, have a wide discretion to reject expert opinion evidence that does satisfy the general acceptance criterion, but is unreliable; see ss135–137 *Evidence Act* (Cth) [76].

The United Kingdom also bases admissibility of expert opinion evidence upon the general acceptance criterion [18]. Furthermore, reliability was held not to constitute a criterion for admissibility [98].

## 4. ESTABLISHING INDIVIDUAL OR SPECIFIC CAUSATION

Supposing now that ($E_{pop}$) is accepted as causal for ($O_{pop}$) in a relevant population. Following Section 2.5, we must now consider whether the exposure is causal of the actual plaintiff's outcome, since in all of the situations under R(1) or R(2), some outcomes in the population are not due to ($E_{pop}$).

When an association between ($E_{pop}$) and ($O_{pop}$) is so established as causal through a test-based statistical procedure such as described in Section 3.4, it can then be used to establish the individual level of the causal chain ($E_{ind}$) → ($O_{ind}$) by the normal legal tests.

In order to avoid complexities, let us for the moment assume that there is one plaintiff who has successfully established that

(a) the plaintiff suffers from the compensable outcome ($O_{ind}$);

(b) there is an exposure ($E_{ind}$) which is sufficiently well defined to enable one to consider whether ($E_{ind}$) actually caused ($O_{ind}$) in the plaintiff;

(c) there is an action (A) performed by the defendant resulting in the exposure ($E_{ind}$) which was in breach of a duty of care owed to the plaintiff.

The legal test to be applied for causation is now prescribed by legislation in Australia. It is provided for under the *Civil Liability Act 2003* (Qld) [75] that for a breach of duty to cause a particular harm, the following elements must be satisfied:

(a) the breach of duty was a necessary condition of the occurrence of the harm ("factual causation");

(b) it is appropriate for the scope of the liability of the person in breach to extend to the harm so caused ("scope of liability").

The scope of liability under (b) is typically a question of law for the judge. We concern ourselves here with factual causation under (a). What constitutes a necessary condition in (a) is dependent upon the nature of the exposure and is considered in Sections 4.1 and 4.2.

### 4.1 Simple Cases—Single Cause

A simple case is akin to the situation R(0) described in Section 2.2, where there is only one possible exposure under consideration. In these cases, factual causation is proven if the exposure is the necessary condition of the outcome. In Australia, *March v Stramare* [96] provided that the approach of the Common Law toward causation in cases involving a single cause of the plaintiff's harm is to apply what is known as the "but for" test.

THE BUT FOR TEST. *"But for" the exposure, would the outcome have occurred on the balance of probabilities?*

It is useful to note that the "But for" test has more than one function. When considering causative issues, the test can also be used in the negative sense. In *March v Stramare* [96] it was used to eliminate, as potential causes, those acts which have no relevance to the occurrence of the plaintiff's harm.

Some American case law has suggested that the balance of probabilities standard for an individual can be satisfied by establishing a relative risk greater than 2.0 in the population from which that individual is drawn. However, there is equal authority for the proposition that a relative risk greater than 2.0 is not a strict requirement of causation; see Carruth and Goldstein [4] for a discussion of the cases supporting the alternative views. *Re Joint Eastern and Southern District Asbestos Litigation* [99] found that if the relative risk failed to reach 2.0, then the epidemiological evidence in combination with other evidence which eliminates confounding factors and strengthens the association specifically in the circumstances surrounding the plaintiff's case, can establish causation.

In Australia, *Seltsam* [100], at paragraph 136, explained that the predominant position seems to be that the balance of probabilities test requires a court to reach a level of actual persuasion, which is not a mechanical application of probabilities. Actual persuasion does not require that a relative risk of 2.0 be reached, even where that is the only evidence put forth. Spigelman, at paragraph 137 [100], speaks of sources of evidence in tortuous claims as "strands in the cable," not "links in the chain." The sources of evidence, being strands in the cable, must be capable of bearing the weight of the inference made, so that one source of evidence alone, or many sources combined, may be capable of establishing causation.

At paragraph 89 [100] Spigelman provided that epidemiological evidence, either alone or in combination with other evidence, is capable of satisfying the balance of probabilities standard and thus establishing specific causation for an individual. His Honour [100] also said at paragraph 29 that further evidence will almost certainly be required where the quality of the epidemiological evidence and the strength of the association are poor. For example, in *Seltsam* [100] it was argued by the plaintiff that asbestos caused renal cell carcinoma even though the relative risk was less than 2.0. It was held [100], at paragraph 171, that it was not enough that the epidemiological evidence only established that asbestos *could* reach the kidney; there needed to be some evidence that the asbestos *did* reach the kidney in this individual case. Relevant medical evidence was required at this step to establish causation: "The strength of the association between asbestos exposure and renal cell carcinoma, and other aspects of the quality of the epidemiological research, particularly inconsistencies amongst the various studies, were relevant considerations which his Honour was obliged to take into account" (at paragraph 32).

### 4.2 Exceptional Cases—Multiple Causes

The "but for" test is not always appropriate, such as in an "exceptional case" in which there are multiple or complex causes of the harm generally operating simultaneously [44] or each capable of causing the harm [5].

The Courts have primarily applied the "material contribution" test to determine factual causation in the exceptional case. Causation is asserted if, on the balance of probabilities, the wrongful act or omission materially contributed to the harm [82]. If the material contribution test is satisfied, then the breach of duty should be accepted as being a necessary condition of the harm and thus the "factual causation" element provided for in s11(1)(a) *CLA* [75] would be satisfied.

MATERIAL CONTRIBUTION TEST. *On the balance of probabilities, did the exposure contribute to the individual's harm to a material extent?*

Determining when the Material Contribution test is satisfied seems to be a difficult question, especially when causation for an individual is being based on results from population studies.

*Bonnington Castings Ltd v Wardlaw* [82] defined material contribution as a contribution which does not fall within the category of *de minimis non curat lex*. That is, for a contribution to be material it must be more than minute or insignificant [47]. *Chappel v Hart* [84] proposes that a mere possibility is not sufficient for causation; rather, the increased risk must cause or materially contribute to the harm actually suffered. Chief Justice Spigelman in *Seltsam Pty Ltd v McGuiness & Anor* [100] went further in describing material contribution as "the 'possibility' or 'risk' that $X$ might cause $Y$ had in fact eventuated, not in the sense that $X$ happened and $Y$ had also happened, but that it was undisputed that $Y$ had happened because of $X$" (at paragraph 119).

Hence, there must be more than a chance that the exposure resulted in the harm; there needs to be some certainty that the harm was the result of the exposure.

With respect to the level of certainty required, *Denzin & Ors v Nutrasweet & Ors* [87] provided that there must be a "probable connection" between the exposure and the harm. Material contribution requires that the exposure must be "the cause" or "a probable cause" of the harm [87]. Therefore, although material contribution may not require that the relative risk in the relevant population be above 2.0, it does seem to require that statistically the exposure significantly increases the risk of harm [87].

So, a mere increase in the risk of harm, as opposed to a statistically significant increase, should then not suffice. However, *McGhee v National Coal Board* [97] substituted "materially increasing the risk" for "material contribution" and this has led to some confusion. In *Hotson* [92], material contribution on the balance of probabilities required that the risk of harm be more than 50%. In *Wilsher* [105], the fact that the court was asked to decide between competing rather than reinforcing risks seemed to be a sticking point even though there were five exposures which all raised the risk of harm, if not materially contributing to it.

The recent decision of *Fairchild v Glenhoven Funeral Services Ltd* [89] has given some clarity to the ambiguities brought about by these cases. *Fairchild* [89] is a case regarding the liability of employers for exposing their employees to asbestos. Because the mechanism by which asbestos fibers precipitate mesothelioma is unknown, the consequences that would follow had it not been for the fault of any one of a number of exposures, such as environmental exposures, is also unknown. Because of the unknown etiology of the harm, the plaintiff cannot prove that the harm was more likely than not caused by the defendant's negligence rather than background exposures. As a result, it is impossible to tell exactly which exposure materially contributed the harm so as to have so caused the harm. Despite the problems posed, the House of Lords found that causation had been established and in doing so gave commentary on past decisions.

The major determination made by the House of Lords in *Fairchild* [89] was to confirm the decision made in *McGhee* [97]. Establishing causation through the simple raising of the risk of harm is to only apply in certain circumstances, so as to work as an exception to the orthodox approach to causation of the exposure materially contributing to the harm. This exception is referred to as the *McGhee/Fairchild* principle. There are no prescribed requirements to be met in order for this principle to be applied. Rather, a case-by-case approach is to be taken, and Stapleton [63] believes that for the time being there will be some uncertainty as to when the *McGhee/Fairchild* principle will be applied.

Nevertheless, from the relevant cases, two important requirements can be established: that the etiology of the outcome is unknown, and that the defendant's conduct had *materially contributed to the risk* that the plaintiff would succumb to the outcome.

There are, of course, more requirements to be met, because otherwise the courts would be subjected to a flood of claims brought under the principle. What constitutes these other requirements is uncertain at present, but some factors emerge.

One such factor clarifies why no causation was established in *Wilsher* [105] despite several exposures materially increasing the risk. This is that there should only be a single type of agent/exposure; as suggested by Stapleton [63], this does not mean that there cannot be more than one agent present, but that agents must all operate in substantially the same way. For example, a person exposed by two neighboring asbestos plants would not be prevented from the application of the *McGhee/Fairchild* principle because the two agents are working in the same way. On the basis of this requirement, the decision in *Wilsher* [105] was correct in not applying the *McGhee/Fairchild* principle. Stapleton [63] provides that in *Wilsher* the evidentiary gap was too wide to leap because there were five agents operating in substantially different manners. The decision in *Wilsher* was not actually creating ambiguity, but rather conforming with the decisions of previous cases; consequently, *Fairchild* approved *Wilsher*.

It is now easy to distinguish *Hotson* [92] from cases such as *McGhee* [97], *Wilsher* [105] and *Fairchild* [89]. *Hotson* provided that exposure upon which the liability is being placed needs to be responsible for more than 50% of the contribution. This only applies, however, if the liability can be quantified with at least some certainty. In contrast, where statistical quantification in the individual is impossible, for example where the etiology is unknown, then all that is required is evidence that the exposure materially increased the risk.

As the facts of the case vary, so do the requirements of the material contribution test. Generally the Material Contribution test will require that the exposure materially contributed to the harm. However, where there is a gap in the evidence such that it is impossible to determine whether the exposure caused the harm, then all that will be required for causation is to prove that the exposure materially increased the risk of the harm.

MATERIAL CONTRIBUTION TEST (where there is an evidentiary gap). *On the balance of probabilities, did the exposure increase the relative risk for the individual to a material extent*?

Where there is an evidentiary gap, the onus on the plaintiff to prove factual causation is less stringent than where there is no gap. Nevertheless, it was expressed in the Ipp Report [37] at paragraph 7.32 that where there is a gap in the evidence, the decision must still be "widely considered to be fair and reasonable."

### 4.3 Allocation of Responsibility

Where there are multiple exposures which may have caused an outcome, and where more than one might be compensable, allocation of responsibility becomes an issue.

This is rarely simple. Consider a plaintiff who has had joint exposure to asbestos and active smoking of tobacco. Population studies [25] show relative risks of lung cancer (O) associated with exposure to asbestos given (approximately) by $(RR_a) = 6.0$, with active smoking by $(RR_s) = 11.0$, and with exposure to both asbestos and smoking by $(RR_{as}) = 51.0$. If all the causal implications are accepted and a relevant subgroup is identified (and we cannot stress too often that these are necessary precursors), then for every *random* 65 persons with outcome O who were jointly exposed, 5 of the outcomes were due to asbestos, 10 were due to cigarettes and 50 were due to the interaction of the two exposures; that is, 55/65 in some sense were contributed to by asbestos. Thus the action of asbestos, for smokers, is to increase the risk far more than it would in the normal population.

How should the courts handle such material contributions, where there is more than one exposure that could be involved in an individual case? The first instance judgment in *Hotson* [92] was to prorate the damages by the assessed risks involved; but this was overturned. On this basis, an interpretation of the "contribution" to the outcome from asbestos exposure would be to decide that all of the asbestos risk, including that for all of the interactively caused outcome, is due to asbestos; that is (ignoring for the moment all other issues such as contributory negligence), since 55/65 leads to a RR greater than 2.0, to award against a defendant who exposed workers to asbestos on a causal argument. But one could argue equally that smoking exposure was responsible for 60/65 of the raised risk in the population.

Legally, the situation may have nothing to do with any comparison of the size of relative risks involved; it may well depend rather on who owes the duty to the plaintiff. An asbestos employer, for example, may have a duty to provide a safe place of work, but must take his employees as they present themselves; and it could then be argued that observation of this duty is (as shown by the relative risks above) rather more vital for smokers than for nonsmokers.

It is of course tempting, nonetheless, to try and find a formula-based method of allocation of the interaction risk based on the two assumed contributors. For example, Chase, Kotin, Crump and Mitchell [7] advocate that the interaction term $RR_{as} = 50$ should be *prorated* in some way. This methodology is superficially attractive, but the actual computations of Chase et al. essentially revert to using an additive model, which in the asbestos/smoking situation (and many others) ignores any synergy involved in the different exposures.

The establishment of an appropriate biological model for the nature of the interaction between multiple exposures is paramount in these circumstances. Epidemiology can play some role in this (see, e.g., [72]), but as Robins and Greenland [50] point out, estimation of these interactions is in general very difficult. Unlike Chase et al., Robins and Greenland conclude that it is the role of the courts, not the statisticians, to decide on the contributions of competing risks in such contexts. Unfortunately, in *McGhee* [97] and *Fairchild* [89] the issue of apportionment was not raised by the defense nor given any consideration by the courts. Hence it remains uncertain as to whether the defendant can be liable for only a proportion of the plaintiff's harm on the basis that there are other exposures contributing to the risk.

Issues of allocation of responsibility also arise where the plaintiff is exposed to a single agent to which several parties have contributed. This is an issue under current debate, but there are several Australian decisions regarding apportionment between tortfeasors. *E M Baldwin & Son Pty Ltd v Plane* [88] approved equal apportionment between tortfeasors. However, this decision has been criticized and not yet been followed. Rather, the courts are taking the view that apportionment does not have to be equal. In *James Hardie & Coy Pty Ltd v Roberts* [94], a recent asbestos related case, Sheller refused to declare equal apportionment where successive employers had exposed the plaintiff to asbestos for different periods, with different intensities of exposure to asbestos with different toxicity. Support of this latter view has been found in several other cases such as *Bitupave Ltd v McMahon* [81] and *Wallaby Grip Ltd v State Rail Authority of NSW* [103].

In the United States radioepidemiologic tables that estimate the probability of cancer developing from a dose of radiation are used to determine the amount of compensation awarded [36]. The probability in the tables is obtained from the assigned share methodology. The assigned share is the ratio of the excess number of cancer cases in the exposed to the total number of cancer cases in the exposed [36]. Furthermore, the population is divided up into subgroups in an attempt to make drawing an inference from the population to the individual theoretically more valid [36]. The assigned share methodology becomes complex and its discussion deserves a paper of its own.

## 5. USING "BARE STATISTICS": THE YELLOW AND BLUE TAXI ARGUMENT

Using statistical arguments for deciding on "balance of probabilities" can sit uncomfortably in the legal situation, despite the similarities in terminology. The use of a relative risk as described above is specifically sanctioned for radiation associated with cancers [27], but in other cases it may well need to be reconciled with legal thinking even if it does in fact describe the way in which balance of probabilities is understood in legal cases.

The confusion about the role of statistics in the establishment of causation is exemplified in the hypothetical case (cited in [92]) of a town having three blue and one yellow taxis operated by different companies. Here, it is postulated that

(i) first, there has been an injury to an individual [the outcome $(O_{ind})$] resulting from being hit by a taxi [the exposure $(E_{ind})$] which was driven negligently [the action (A)]; and

(ii) there is no evidence in the accepted sense of which taxi company is responsible for the action.

One might argue that, in the light of no other evidence, there is a "75% probability" (i.e., a 3 to 1 chance) that the taxi is blue, and therefore based on the "balance of probabilities" the action, and hence exposure and outcome, were caused by the blue taxi cab company.

Such reasoning using "bare" statistics was disallowed by Lord Mackay in [92], largely on the grounds of the perceived inequity to the defendant, who would have to bear the liability for all cases; this seems unfair even if they are indeed responsible for 75% of them.

How is this argument resolved in the light of our proposed procedures for asserting causation? If we accept the proposition that a relative risk greater than 2.0 establishes causation on the balance of probabilities, then we might appear on the face of it to be exactly in the situation of the blue and yellow taxis.

To resolve this, we examine the causal links in the taxi case more carefully. The link $(E_{ind}) \to (O_{ind})$ is well established as necessary and sufficient, fitting into the category R(0) of Section 2.2. It is the first link $(A_{ind}) \to (O_{ind})$ which must be established. If we are to ascribe the probability of the outcome for the plaintiff in the same 3:1 ratio proportional to the number of taxis, there are two crucial assumptions that we need to make, both of which must be tested from the nontechnical or commonsense viewpoint. These are that:

(a) the probability of the action (negligent driving) is the same for each taxi;
(b) the probability of the exposure (being hit) from each taxi is the same for the individual.

Under (a) any taxi is as likely as any other to hit the plaintiff; under (b) the taxis to which the plaintiff is exposed are three times as likely to be blue. This will then give a RR greater than 2.0 of being hit by a blue taxi, and hence give a real "balance of probabilities" that the blue taxi has caused the accident, analogous to the epidemiological concept of risk.

It is (b) that essentially says that we need to have a *population risk estimate* to add to the "naked statistic" of the chance of being exposed. Possibly the only way to verify this is to use population-based data to assure that all taxis are equally likely to be in the neighborhood of any given person in the relevant population, and that the taxis are all equally likely to hit people in their vicinity. In this case in fact it is an estimate of $(A) \to (E)$ that is in question, rather than $(E) \to (O)$, but the principle is the same. In contrast, the initial argument merely told us about the chance of picking a taxi at random without consideration of the causal links and assumptions involved; it did not tell us about the chance that an injury was caused by taking this cause proportional to its true population level of risk.

Furthermore, the hypothesis that blue taxis cause three times as many accidents in the population, if subjected to the scientific methodology proposed above, fails many other population-based tests for establishing causality. There is no information given about Tests 4 to 9 described in the Appendix, in particular covering issues of validity of data, other confounders, repetition, exposure-response and statistical significance. In effect, there is no underlying proof of $(A_{pop}) \to (O_{pop})$ and no corresponding information about $(A_{ind}) \to (O_{ind})$, and this seems to be why scientific methodology, the legal reasoning of Lord Mackay in [92] and common sense all stand together to reject the hypothetical claim.

This is still not quite analogous to using epidemiological risks. It may be helpful to extend the description to a more relevant one. Suppose that statistics on traffic accidents caused by taxis had been kept for, say, twelve months, and that for each such accident we had recorded the color of the taxi. Furthermore, suppose that in this situation, three times as many people had indeed been hit through negligent driving by the blue taxis compared with the yellow taxi.

Are we then prepared to accept that the blue taxis are, on the balance of probabilities, responsible for the single accident we have observed?

We need to make several assumptions again for this to be valid. We need (a) above, still. For suppose, to the contrary, that by a licensing arrangement the only taxis allowed in this area were yellow; obviously we would change our views. We need to be satisfied that the same drivers, or mechanics, were still driving and maintaining the taxis so that circumstances in our case were still similar to those where the data had been collected, and that the population was still relevant to the plaintiff. Would the courts now find against the blue taxi cab company?

Perhaps not. Certainly, in the absence of such external evidence, both the assumptions (a) and (b) seem inherently unreasonable and indeed manifestly unfair. But regardless of their equity, the underlying *statistical* reason to reject the initial argument is because it is based on the wrong set of probabilities: it argues on the probability of the taxis being chosen at random, rather than on the relative occurrence of actions and exposures leading to outcomes, which (without these assumptions) may be totally different.

Thus our suggested use of relative risks in Section 4 is a different situation to that of the taxis, and it appears to encapsulate the question of arguing from the general population to individual balances of probability.

## 6. CONCLUSIONS

There are marked similarities between statistical and legal thinking concerning the onus of proof in establishing causality of a relationship.

In statistical or scientific arenas, the onus of proof is essentially on the scientist to show that a relationship between exposure and outcome is causal. The initial working assumption in the scientific method is indeed that no relationship, much less a causal relationship, holds. Unless the data, the theories and the inferences disprove this assumption and thereby establish causality, then the initial assumption remains unrefuted.

At the population level, we have given one set of tests for establishing (at least at a working level) a standard of scientific proof which might be accepted in legal cases in discharging this onus of proof.

This approach is often made quite explicit in common systems of statistical inference, where a so-called null hypothesis, that there is no relationship (causal or otherwise) between the exposure and the outcomes, is actually formalized. It is precisely because this must be falsified that the test for statistical significance is set in the way it is, especially in comparison to balance of probability arguments. A statistician is not convinced of the existence of a relationship by the mere 50% chance that the result is not random; by convention a more than 95% chance of the outcome being unexplained by random events, or a more stringent level of, say, 99% or 99.9% (or even in some cases a more lenient level of, say, 80%) is required before it is accepted that this aspect of the onus of proof has been discharged.

However, mere discounting of random events as an alternative explanation by demonstrating statistical significance is only one (albeit a central) part of the test-based approach. Discounting other possible explanations is also critical.

We have argued that these scientific tests are crucial, but incomplete input to any legal tests for causation. In terms of the causal chain, they may establish the fact that population risk of the outcome in question is causally raised by exposure; the other input needed is a method of arguing from the causation of risk for a population to causation of actual outcome for an individual.

We have considered whether the term "balance of probabilities" can fruitfully be equated with the scientific calculation of "relative risk greater than 2.0," based on the establishment of a relevant population, the use of the estimate of the lower confidence bound as a more stringent estimate of the excess risk in such cases, and the role of other evidence in providing this individual risk.

This is not to be confused with the different approach that might be taken for policy decisions. Because of the potential harm from waiting for associations to be scientifically proven, it can be argued that in implementing a policy of prudent avoidance of suspected but unproven risks, the onus of proof should at times be reversed [17]. Thus if there is either a theory, or a set of preliminary data, suggesting a harmful relationship by, for example, showing a value of $RR > 1$, it may be appropriate in that arena to initially accept this relationship and act accordingly, and to abandon it only when or if the relationship is falsified. There have been some notable examples (the relationship of caffeine and pancreatic cancer [41] or the relationship of cadmium and prostate cancer [11]) where initial indications were overturned as more studies were carried out. The full range of tests in Section 3 above would not have yielded a proof of causation in such studies.

Similarly, in the legal context the onus of proof is supposed to be on the plaintiff. This is certainly the basis for the decision in *Wilsher* [105], in which it was held that it was not the task of the defense to show that one of the other risk factors was responsible, but rather the task of the prosecution to show that the defendant was responsible. Legislation in Australia has provided for this (see in particular s12 *CLA* [75]), but *Shorey v PT Ltd* [101] has suggested that the onus can shift to the defendant to disprove the plaintiff's case. Nevertheless, the plaintiff must prove causation in the relevant population, by first passing the tests described in Section 3 to a satisfactory level. To obtain a judgment on the balance of probabilities on an *individual,* the plaintiff must then show that the population rates are applicable to the individual, either for $E^C$ alone (in which case the *but for* test is passed) or over a set of exposures to which $E^C$ has a material contribution (in which case the test of *material contribution* is passed). Nonetheless, a raised RR may still not establish causation in the legal sense if common sense suggests otherwise, or in the statistical sense as discussed in Section 4.

Given all of these issues, it is perhaps inevitable that the statistical viewpoint will often aid the defendant more than the plaintiff. It can aid the court in understanding the validity or invalidity of a statistical argument, the support or otherwise of data for a particular case, the responsibility of randomness of observational studies for observed population results, and the identification of individual risks and assessment of risk for subgroups with similar characteristics to the individual in question. Good statistical analysis should give the court a better idea of the risks in the population.

Although all the contributing exposures, including nontortious background exposures, are relevant when considering causation, there is somewhat of an exception to this. A principle of law, the Eggshell Skull Principle, provides that a victim must be taken as he or she is found by the negligent defendant. Consequently, if the plaintiff is more susceptible to harm because of an inherent weakness or disorder, the defendant cannot argue that the exposure was trivial and that it would be unfair to burden him or her with the liability, nor does it render the damage unforeseeable [101]. The same onus of proof applies such that the plaintiff must prove that the defendant's negligence caused the harm. Causation is proven in the same fashion as in any other negligence claim. In *Shorey v PT Ltd* [101] it was provided that if the defendant is to escape liability, he or she must prove that another causative exposure had taken over as the effective cause of the plaintiff's damage.

Statistical analysis can never identify the cause of a single plaintiff's outcome *unerringly*; it cannot help if a court uses Stapleton's standard of "... the production of a latent bodily condition *certain* to produce disabling personal injuries in the future" [62]. Nor can statistical analysis generally help with the question of negligence, that is, the step (A) → (E) in our causal chain, but the introduction of the *CLA* [75] may lead to statistics having a part to play in this step. Before the recent reforms, the risk of harm being found by the courts as reasonably foreseeable (i.e., the question of whether an action was negligent) was a certainty. Now, s9(1)(b) provides that the risk of harm must also be "not insignificant." The Ipp Report [37], whose recommendations were the basis for the *CLA* [75], provided that:

> *"The phrase 'not insignificant' is intended to indicate a risk that is of a higher probability than is indicated by the phrase 'not far-fetched or fanciful', but not so high as might be indicated by a phrase such as 'a substantial risk'. The choice of a double negative is deliberate. We do not intend the phrase to be a synonym for 'significant'. 'Significant' is apt to indicate a higher degree of probability than we intend."* – at paragraph 7.15.

This test is not to be used when considering causation of harm nor whether the kind of harm was reasonably foreseeable from the kind of exposure. It is only to be used to determine whether an act was negligent.

Statistics may be used to determine whether a risk is not insignificant, but this has not yet been seen in the courts. Currently it is only if the court accepts the step from population risk to individual cause on a balance of probabilities, or accepts a risk estimate in deciding on whether there has been material contribution on the balance of probabilities, that sound statistical methodology can assist in moving from the use of "bare statistics" to providing a sound underpinning of decisions made in these most difficult and nontraditional problems. Even then the statistical evidence may not be strong enough on its own to prove causation, but when used in structured combination with other evidence it can create a very convincing argument for causation.

# APPENDIX A: A TEST-BASED APPROACH TO CAUSALITY OF RELATIONSHIPS

Test-based approaches to causality in epidemiology date essentially to the work of Bradford Hill (see [30, 31]). Although there are many aspects of his set of tests which can be criticized [66, 70], in establishing the population-based relationship $(E_{pop}) \to (O_{pop})$, for use in a legal context we feel such an approach is still appropriate. Here we spell out and illustrate ten such tests, grouped into four types: theoretical, empirical, statistical and inferential.

## A.1 The Theoretical Step

The first set of tests for a causal association relates to the need for an underlying theory.

TEST 1 (Existence of mechanism). *Is the proposed association explained by a biologically plausible mechanism?*

Surprisingly, although passing Test 1 may seem to a layman to be of paramount importance, it is certainly not always regarded as vital that it be passed. For example, we are still at an early stage of understanding cell-level biology; the mechanism for causing various cancers is quite unknown and therefore we cannot and do not expect that the role of an individual exposure will be explicable in detail. Thus in the case of exposures suspected of causing cancer, such as environmental tobacco smoke or asbestos fibers, we do not have any established way of passing Test 1. Also, it is not always a reasonable question given current scientific knowledge. Rothman and Greenland [56] recognize that biologic plausibility may be solely based on prior beliefs and not data or logic.

However, one cannot underestimate the value of a real and positive answer to Test 1 in establishing causality. In revealing the role of sexual activity in causing the spread of AIDS, the existence of a biological mechanism was paramount: locating an AIDS-causing virus that could be spread by such activity gives a much stronger proof of the causality than is available for a theory of causation by use of amyl nitrate, where no such explanation was found.

There are some cautionary notes in assessing Test 1.

It is clearly desirable that the theoretical basis be verifiable in some way independently of the population data collected and used in the steps below, but this is not always the case. Given the empirical observations, it is quite common and accepted scientific practice to develop *models* for the biological processes involved, based purely on describing the observations rather than on building a biologically "causal" model.

These may have an underlying biological rationale or they might be purely descriptive of the data, such as the model of Doll and Peto [13], $RR(x) = (1+x)z$, where $RR(x)$ is the relative risk of lung cancer associated with smoking $x$ cigarettes/day, compared with not smoking. This may look to a nonscientist like a biologically based explanation of the increase in relative risk, when in fact it is purely a convenient mathematical way to describe data. In contrast, theoretical models of the structure of the cancer growth have been developed which can then be assessed for goodness of fit to the data.

In this vein one may look also at the IUD/PID relationship. There is an established theory of this mechanism for, say, the Dalkon Shield: the long "tail" of the device is supposedly a conductor of infection, and the lack of such a tail is the reason advanced for lesser associations with other devices. This is plausible; but is it a theory developed after the data or a theory with a sound and commonly agreed backing from biological argument?

Despite the problems involved, biological knowledge should not be discounted; it should just be recognized as difficult to apply. In any event, if Test 1 is failed, one often resorts to its weaker cousin, Test 2.

TEST 2 (Analogous relationships). *Is the proposed association analogous to some other accepted causal association?*

Test 2 is often used as a surrogate for Test 1. For example, if it is accepted that active smoking causes lung cancer, then hypothesis that exposure to environmental tobacco smoke (ETS) causes lung cancer might be posed and Test 2 accepted by using the analogy with active smoking; indeed, exactly this argument was advanced by the U.S. Environmental Protection Agency in its conclusion that ETS is a carcinogen [15].

However, care needs to be taken when considering analogous relationships, because scientists can use their imagination to find analogies to any accepted mechanism (see [57]). It is a matter for subject area expertise, not for statistical expertise, to decide how strong such analogies are. The ETS analogy must be based on a decision that any toxicological, site, or other differences between the types of exposure are irrelevant. A similar analogy between lung cancer effects of exposure to forms of industrial asbestos, say chrysotile and crocodilite as mined at Wittenoom, Australia, may be valid, but this also needs to be ascertained.

We believe that in many cases the role of Test 2 is unfortunately confused, especially when it supplants rather than separately reinforces Test 1. It is clearly of great value in *initiating* an area of study, but following the central dictum of science that one cannot prove something merely by asserting it might be true, it seems that Test 2 on its own is a weak addition to establishing a causal inference.

As an example, consider again the Wittenoom asbestos example. If one accepts the chrysotile exposure results of [1], then Test 2 indicates that exposure to crocodilite may be causal for lung cancer. One then initiates a study of, say, those exposed to crocodilite at Wittenoom. If this study enables further tests such as those below to be passed, the analogy argument certainly may appear to have greater validity; but if they are failed, then one should see the analogy argument as being falsified, rather than as continuing to have self-sustained force.

In some studies, even failing Test 2 is not seen as a major problem. In the study of poppers and AIDS this was the case: here the level of theoretical proof appears to be at the weakest form of Test 2, namely that amyl nitrate is a "foreign substance" to the body and therefore might by analogy with other foreign substances be accused of causing almost anything.

Regrettably, too often the use of these tests is at this weak level and we are forced to ignore this lacuna in the argument for causality. And yet, as Sir David Cox reinforces [10], without a biologically or physically plausible model the leap from association to causation is a much less convincing one.

Finally, one must point out that due care has to be taken in separating the "hypothesis-generating" information in analogous studies from the "hypothesis-confirming" steps below. Sir Richard Doll [11] notes an occasion on which this error seems to have led to an incorrect conclusion that cadmium had contributed to the development of prostate cancer. When the hypothesis-generating studies are purely comparative or anecdotal, this may not be a problem; when, as with the exposure to spousal smoking, they seem more rigorous (e.g., the cohort study of [33]), then the temptation to include them in formal analysis can be strong.

The third test in this theoretical step is that of temporality.

TEST 3 (Temporality). *Does the exposure precede the outcome?*

Little needs to be said about this in theory: if it fails, then the hypothesis of causality can be eliminated, but if it is passed, we have little further support for the hypothesis, since simple precedence, as pointed out by philosophers at length, does not imply causation (see [55]).

Moreover, even where this test is failed and the outcome does precede the exposure, it does not necessarily follow that the exposure does not cause the outcome; it only shows that the exposure could not have caused the outcome in the particular circumstances present; see [57] for further discussion.

We note, however, that this test is often hard to apply in the epidemiological arena. The time of actual onset of a disease, for example, is frequently indeterminate, and for exposures such as environmental tobacco smoke, it may be difficult to tell whether Test 3 is passed or not since the definition of the exposure is not simple and proper account of latency time may be difficult. As another example, in assessing the association of an IUD with pelvic inflammatory disease, the time of exposure may be much better established than is time of exposure to ETS, but time of disease onset (unless acute) may be at least as hard to establish as for lung cancer.

If passing this test were seen as critical, then on occasions it would be almost impossible to establish causality in population-based studies. In general, when the relevant times are hard to establish as above, the test is essentially overlooked. However, it can sometimes be valuable in falsifying a causal hypothesis; for example, this test was used in showing that stress in pregnancy does *not* cause Down's syndrome, since the physical changes start earlier than the empirically measured stresses in the studies claiming to provide proof.

### A.2 The Empirical Step

Whether because of a firm theory or because of analogy, the next step in attempting to establish causation in observational studies is the *empirical* step: collecting data which relates to the asserted causal link and considering the degree of support given to the assertion by such data.

Consider again the measure of association in a given empirical study to be the *relative risk*, denoted RR. In theory this describes the expected rate of outcome per exposed person divided by the expected rate of outcome per unexposed person.

There are also inherent complexities in measuring such risks; for more details see [50], who point out the extra problems which nonstandardization and interpretation may cause.

The first test in the empirical step is too often overlooked in scientific analysis, although perhaps it is overemphasized in legal contexts. It is certainly supplemental to the usual Bradford Hill tests for causality.

TEST 4 (Validity of data). *Are the data collected valid?*

There are a number of aspects relevant to answering this question. Feinstein [16] raises many issues which need to be clarified in nonrandomized studies, as do Chalmers [6] and Mosteller and Chalmers [46].

First, and regrettably, it must be admitted that all too often arguments are based on data containing mistakes. It is very easy for such errors to creep into studies, and forgiveness of human error may be partly why this step is overlooked in scientific argument. In practice, a small number of mistakes in data (especially in a large study) should not affect conclusions if they are otherwise clear-cut. When they are not, the effect of the errors may be substantial.

For example, the data by Hirayama [32, 33], on which one of the first assertions of a relationship between exposure to ETS and lung cancer was made, appears to contain some subjects who are dead in 1981 and alive again in 1984, although this did not substantially alter the conclusions. Garfinkel, Auerbach and Joubert [19] showed that classifying smokers as nonsmokers may give a 25% invalidity rate in hospital studies of this same phenomenon. In this example, the empirical relative risks are only around $RR$ from 0.8 to 1.5 [67], so the effect of such errors may be substantial: on the basis of data available at the time, Lee [39, 40] held that misclassification of subjects could account for the whole of the observed raised risk of lung cancer in ETS-exposed subjects. Another dramatic example is given by Kronmal, Whitney and Mumford [35], who claim that much of the data on which the Dalkon Shield was evaluated might have been in error in substantial ways, and that up to one third of these data may have been omitted from analysis; the reanalysis in [35] claims that the original study overstated the relative risk (originally set at $RR = 12.0$) by a factor of some 30%. This issue is also addressed by Lee, Rubin and Borucki [38]. See also the discussion in [24] and [52].

Second, studies may be prone to different *systematic errors and biases.* These include biases due to interviewer practices, questionnaire design, poor handling of missing values, poor definitions or measurement practices, or lack of representativeness of the study; the list is long. Feinstein [16] has a very good discussion of these issues. But if these biases exist, then the validity of the data is impaired and the conclusions cannot be automatically trusted.

In passing or failing this test one needs, in general, far more information than is usually available to anyone other than the original researcher. Usually, at most an outsider can use general statistical principles to identify the potential problems of a poor study using tests of principles of design. It is often only when access is given to raw data (as in the Kronmal, Whitney and Mumford [35] reanalysis of the Dalkon Shield and related data), or occasionally when dual or followup publication occurs (as in the Hirayama study [32, 33]), that data-based critical review is possible.

Because the detection of errors or design flaws is much easier to raise in nontechnical ways, it is often part of the normal legal procedure to address such questions at some length. Statisticians can play a useful and constructive role in this by developing "what-if" approaches to see how badly a study might be affected by potential biases and errors (see [67, 69] for a relevant example). This enables one to assess whether the degree of failure on any of these matters of fact is sufficient to invalidate the conclusions drawn or just weaken their strength.

Now let us turn to the analysis of studies assuming they are validly conducted. There are then a number of further tests which can be considered in this empirical step.

TEST 5 (Strength of association). *Is the observed association a strong one*?

There is a subjective assessment needed to decide that an association, measured for example by a relative risk, is strong. How strong is strong? All that can be said rigorously is that the contribution to proof of causality is greater for larger observed relative risks (and even this has to be qualified as in the discussion of the statistical step below). Indeed, Doll [11] downplays this aspect of testing for causality to almost a nonissue.

An ancillary reason for preferring to accept stronger associations is that they also allow for a little laxity in Test 4. Where there is strong association between the exposure and the harm, it is not likely that this association could be explained by another factor that was not considered, that is, an omitted variable (OV). In order for the OV to explain the association between the exposure and the harm, two conditions must be met. First, the relative risk of the OV must be greater than that of the requisite exposure. Second, the prevalence of the OV in the exposed group must generally be substantially greater than that in the unexposed group. This is known as the Cornfield Inequality [9] and is discussed by Yu and Gastwirth [74].

There is some agreement that a relative risk below 2.0 is not indicative of a strong association [11, 43] and may occur through misclassification and other errors. Associations with such low relative risks are only likely to be accepted as causal if the other tests are passed at a much more stringent level. It is much harder to observe spuriously generated relative risks above the level of, say, 3.0 or 4.0, even from studies with systematic biases or errors in data.

It must be stressed that, although a positive answer to Test 5 is a useful part of building the case for causality, a negative answer is not of much use in demolishing such a case. A weak association does not mean that there is not any causal connection. There may well be exposures which cause cancers, say, but only contribute a small fraction of the overall population rate of occurrence; if studied in sufficient detail, these could lead to weak but still accurate measures of association.

A strong association between the exposure and the outcome merely eliminates the possibility that a weak confounder or some other bias is entirely responsible for the association.

TEST 6 (Lack of confounders). *Are there other aspects of the study group that might explain the observed association*?

The need to establish a causal link for the individual certainly appears to require that (to some appropriate and reasonable extent) there are no other potential causes which may have led to the observed outcomes. Thus removal of confounding effects must be a key consideration when invoking results of epidemiological studies in legal situations, but it is unrealistic to believe that all such effects can be removed; see [12] and [56]. This necessitates quite explicit evaluation of the components (A), (E) and (O) of the chain to uncover the real links in the tricomponent model described above.

Examples of confounders and their interaction are given in Section 2.4. For example, the purported exposure (E) may be indirectly associated with (O) through an independent exposure in such a way that there is in fact no causal relationship between (E) and (O). The problem becomes more complex when a number of different exposures are involved, in which case they may impact on (O) independently (and hence conform to an additive model of risk) or interactively (in which case a synergistic description may apply). Disentangling the links is vital to understanding the impact of exposures on outcome and hence whether and to what extent (E) can be held responsible for (O).

Alternatively, it is also possible that the risk factor lowers the estimated relative risk so as to mask a meaningful effect. This occurs where the risk factor has a relative risk greater than that of the exposure in question and is more prevalent in the unexposed than the exposed group. This is known as the Reverse Cornfield Inequality; as Yu and Gastwirth [74] write, it is especially useful when a study yields a "suggestive finding" such as a RR of 1.50.

Where the epidemiologist has an understanding of the possible biases so as to be able to give a quantitative assessment of the bias, the uncertainty can be accounted for by using a sensitivity analysis to illustrate the possible extent of the biases. However, when the possible cause of the bias is unknown or there are several different exposures, then an ordinary sensitivity analysis should be abandoned. The accuracy in the quantification of confounding effects can be improved upon by using other methods such as Monte Carlo sensitivity analysis and Bayesian bias analysis (see [64]).

Thus this test, although part of the empirical step, is also one in which statistical expertise is of value. Furthermore, part of the professional skill of a statistician lies in designing studies without such confounding factors, or in designing forms of analysis which can take them into account.

The next test is in many ways the key to establishing scientific proof in a traditional sense. It requires that, for a causal association to be established, the same effect should follow the same cause in a number of studies, or in repeated studies, in a consistent way.

TEST 7 (Consistency of association). *Is the association consistently found over a number of studies?*

There are a number of ill-defined terms in this test. How consistent is consistent? How many studies are needed? Must they all be the same in general structure?

As so often, these are a matter for judgment. But what is clear is that an association observed only once, no matter how clear-cut it seems, should not be taken as causal unless it can be repeated. Of course, the stronger and more clear-cut the first observation, the more likely we are to believe that we have "almost" established causality; but then, if the causality is so obvious, often the construction of confirmatory studies should be almost trivial. Of course, with a large positive study there is a tendency to expect that further studies are not needed; but we again cite Doll's example [11] of cadmium and prostate cancer as an illustration of the caution.

It should be noted that consistency of results among studies does not necessarily mean that their statistical significance is the same. Nor for that matter are studies with differing significance results inconsistent. As has already been mentioned, there is almost always more than one causal factor present. Hence, the significance of association for the main exposure may vary between populations and times because the other causal factors will not vary in the same way as the main exposure, as suggested by Rothman and Greenland [57].

One aspect of this test which appears worth stressing is that if all other factors are equal, later studies should in principle provide better confirmation than earlier studies. This is not only because of the simple point that the early studies are often, inherently and unavoidably, hypothesis generating. It is due to the more subtle fact that if a relationship is indeed causal, then later studies should be designed to avoid confounders and other pitfalls to which earlier studies are prone. Hence causal associations should be more and more consistently visible as time progresses. However, this may be mitigated by "remediation bias," in which later studies may not reproduce earlier stronger results because of actions, such as reduction of exposure, taken on the basis of the earlier studies.

There may be exceptions to this test. If an association is geographically or temporally specific, then extra studies elsewhere or at another time may not confirm it, but then as with any confounding we do not have a general relationship anyway and only the relevant population must be considered. If the biological mechanism is well accepted, we may relax the need for extra studies, for example in the case of exposure to nuclear explosion radiation and outcome of cancer. But in general, repeatability, both conceptually and in fact, lies at the basis of the scientific method and one would need strong reasons not to require it as part of a proof of causality in population studies.

### A.3 The Statistical Step

The next step in establishing a causal association is the statistical step; this will most often in practice take place simultaneously with the empirical step.

As detailed above, in Tests 4–7, we have been acting as though there were a true measured association, known exactly, whose strength we could assess and whose consistency we could judge.

This is not true. Even in a situation where our empirical studies are well designed, there is still the role of chance (or luck, as Doll and Peto call it [13]) in allocating the characteristics of the actual group of individuals studied.

Hence, we may observe a high relative risk when in fact there is no effect at all, just because by pure bad luck we saw far more effects in the particular (usually small) group of exposed subjects examined than we should have; conversely, we may miss a true association (especially a weak one) for the same reason.

Thus we use the concepts of "statistical significance," "confidence intervals" and "power" in evaluating the validity of associations observed. Around a relative risk of RR we may construct a so-called confidence interval (CI), comprising an upper confidence limit (UCL) and a lower confidence limit (LCL). This gives a range within which we are "confident" that the true measure of association lies; the technical interpretation of "confidence" in this context is important and differs in frequentist and Bayesian paradigms.

Although a 95% CI is often adopted, confidence intervals of different size (e.g., 90% or 80%) are also used in certain situations, depending on the degree to which one wishes to rule out other possible explanations, in particular chance.

The first consequence of this approach is to give us a *range* of values for which we should use the tests of the previous section. For Test 5, on Strength of Association, we should thus perhaps be more stringent for the type of compensable cases considered here: we really want to know that the LCL rather than the relative risk RR itself is large, since a relative risk of 10.0 is not of great value in asserting a strong association if the LCL is only 0.5! Doll [11] cites a value of 2.0 for the LCL as one where he feels there would be "... seldom any difficulty in accepting a hazard."

Conversely, for Test 7, on Consistency of Association, we can be somewhat more lenient, once we recognize that there is uncertainty in the estimation of the true relative risks and we can compare the ranges (LCL, UCL) compatible with the studies to see whether there is consistency or not. In addition, many meta-analysis techniques which are used to combine results from different studies (see [28]) may enable us to identify heterogeneity, or extra unexplained inconsistency, between studies.

Perhaps more central in the use of the statistical step, and certainly the most obvious use of statistical thinking in most presentations, is the idea of statistical significance. If in fact we had no association, then the true value of RR should be 1.0. But if 1.0 lies outside the range (LCL,UCL), it follows logically that there is a less than 5% chance (i.e., less than 1 in 20 chance) that the association is due to a random choice of subjects (i.e., "bad luck"), and then we say the relative risk, or the association itself, is "statistically significant."

This sketch of statistical practice is intended to assist the interpretation of the crucial next test.

TEST 8 (Statistical significance). *Is the observed association statistically significant?*

If not, then in the same way as in Test 6, we have failed to exclude one of the possible causes of our observations: in this case, not a physical confounder but the confounding effect of sheer variation in our choice of subjects, that is, the effect of chance.

In practice, for compensable cases this test should precede many if not all of the tests in the previous section. As Doll points out [11], "that chance may be the explanation of a raised [risk] is, of course, the first possibility that any epidemiologist will consider." For if this test is failed (i.e., if the results are not statistically significant), then the questions of the strength of the association, or whether the association might be caused by confounding factors, or even questions such as temporal ordering or biological plausibility, are moot at least until more information is gathered. We have, if Test 8 is failed, a phenomenon that could have been caused by random fluctuations in the population alone.

Thus we certainly cannot describe a causal relationship as proved.

Failing this test certainly does not show the causal relationship is false. Lack of significance is often no more than an indication that the study is too small to prove the hypothesis; one toss of a coin, for example, can never convince us that we have a double-headed penny. Alternatively, if an outcome is sufficiently rare in a population, then even a large cohort study may fail to have the power to detect it. A meta-analysis also makes it possible to see whether, in combination, chance is excluded in the pattern of observed results even if the individual studies are not significant; if used with care, this can be a useful tool in the assessment of significance.

TEST 9 (Dose-response relationship). *Is there an increase in magnitude of outcome from an increasing level of exposure?*

There is a second, essentially statistical, test which if passed is an extra building block in constructing a proof of the causality hypothesis.

This is often referred to as the "coherency" test. This test will be passed if there is cause-and-effect interpretation from the epidemiology study which is consistent with the history and biology of the disease. There will be times when this test is failed: if, for example, we have a lethal substance at any level of exposure, or if there is a threshold effect and all observations are above the threshold. In general, however, a biological or physical plausibility argument is used to justify the use of this test, and it is certainly a convincing test when passed.

Test 9 also interacts with Test 4. If there is a systematic bias in the data (say, systematic misclassification, as is possible in the ETS example [67]), then we might not expect such bias to increase with increasing exposure; so even if we had established significance under Test 8 we would find that Test 9, for dose-response, failed. This may indicate, for exposures not instantly toxic, that although there is some nonchance exposure producing the result we observed, it might not be the hypothesized causal agent, but may be some other agent entirely. Examination of possible confounders, as described by Test 6, may help to explain this. Alternatively, associations that have a dose-response relationship may not necessarily be causal because a confounding effect may demonstrate a biological gradient in its relation with the disease, as provided by Rothman and Greenland [57].

Why is this a statistical test? In order to ascertain if there is a real increase in outcome (i.e., increase in risk), not just a random fluctuation at different dosage levels, we need statistical techniques similar to but more sophisticated than those which allow us to decide that the overall effect is statistically significant. These often rely on models of some complexity: linear or log-linear models, often derived for mathematical convenience with little or no biological underpinning, as discussed in Section A.1. As mentioned in Section 4, in *Seltsam* [100] Spigelman did consider dose-response relationship as a relevant factor; however, because of the more technical nature of this test, there may be inappropriate and unquestioning acceptance of a study's reported dose-response results in the legal system, and it can be hard to get sufficient data to verify that the test does hold. Nevertheless, the effort to address this test is worthwhile, and indeed Doll [11] stresses the value of coherency in responding not just to intensity of exposure but also to duration of exposure, and even to time since first exposure.

More than most, this test may be failed without too much penalty; in many cases there is just not enough data to make clearcut judgments, as is the case in the ETS and lung cancer example [68].

### A.4 The Inferential Step

There is a final step if all or some of the above tests have been satisfied. This is the inferential step, in which the individual pieces of the argument are joined together to form a logical conclusion.

TEST 10 (Validity of logic). *Is the conclusion actually justified by the data and analysis presented?*

This seems trite, stated in this way. And yet we find in, say, Wu-Williams et al. [73] the statement that "ETS is causal for cancer despite the data shown here."

Such a test should, ideally, involve attention to all of the preceding tests, and if a causal relationship is claimed, then the degree to which it is based on each test should be defined properly. This is rarely if ever done in a formalized sense; indeed, even the Bradford Hill criteria, which are less explicit than those posed here, are rarely "ticked off" in any systematic way. For example, the second EPA Draft Report [15] (see also [14]) tests the causality of ETS for lung cancer by using, quite explicitly, only Test 2 under the claim that this test is so well satisfied that no other approach is needed. There are indeed celebrated instances in which a systematic approach of any kind seems sadly missing: Stolley [65] opines that R. A. Fisher, in defending no causal link between active smoking and lung cancer, had used "incomplete and highly selected data... with scant attempts to weigh the evidence or reveal the obvious deficiencies in his data." On the other hand, even when there are attempts to validate the tests above (or versions of them), it is often the case that Test 10 is only weakly satisfied.

So the final test is not always as simple-minded as it may seem and, overall, it is the most important: as Doll [11] states, that "if there is one general rule, it is that conclusions can be reached only after the totality of the evidence is taken into account."

Scientific proof of causality may always be fragile, always falsifiable. For some reasons, such as taking precautionary steps in setting public health policies, it may not be necessary to prove an association is causal in this way.

But in the legal context, in which a court will require proof that an exposure (E) actually is causally related to an outcome (O), it is vital in our view that at least the pragmatic approach detailed here, culminating in a firm and logical statement of the passing of tests, should be required before any such conclusion is attempted.

## ACKNOWLEDGMENTS

This paper benefited by early discussions on the statistical aspects with Geoff Eagleson, John Eccleston and David Scott, and by the invaluable advice on some legal aspects of Glenn Eggleston and Richard Travers, to whom we are very much indebted.